\documentclass[%
 reprint,
 amsmath,amssymb,
 aps,
]{revtex4-2}

\usepackage{graphicx, amsmath}
\usepackage{epstopdf, epsfig, subcaption}

\begin{document}

\title{Closures for Multi-Component Reacting Flows \\ based on Dispersion Analysis}

\author{Omkar B. Shende}
    \email{oshende@stanford.edu}

\author{Ali Mani}
    \email{alimani@stanford.edu}

\affiliation{Department of Mechanical Engineering, Stanford University, Stanford, CA 94035, USA}

\begin{abstract}

This work presents algebraic closure models associated with advective transport and nonlinear reactions in a Reynolds-averaged Navier-Stokes context for a system of species subject to binary reactions and transport by advection and diffusion. Expanding upon analysis originally developed for non-reactive transport in the context of Taylor dispersion of scalars, this work extends the modified gradient diffusion model explicated by Peters (Turbulent Combustion, 2000) and based on work by Corrsin (JFM, vol. 11, p.407-416) beyond single-component transport phenomena and involving nonlinear reactions. The presented model forms, from this weakly-nonlinear extension of the original dispersion theory, lead to an analytic expression for the eddy diffusivity matrix that explicitly captures the influence of the reaction kinetics on the closure operators. Furthermore, we demonstrate that the derived model form directly translates between flow topologies through a priori and a posteriori testing of a binary species system subject to homogeneous isotropic turbulence. Using two- and three-dimensional direct numerical simulations involving laminar and turbulent flows, it is shown that this framework improves prediction of mean quantities compared to previous results. Lastly, the presented model form, collapses to the earlier gradient diffusion and its modified version derived by Corrsin in the limits of non-reactive species and linear reactions, respectively. 

\end{abstract}

\maketitle
\onecolumngrid

\section{Introduction}
Building reduced-order models for turbulent reacting flows is theoretically and computationally challenging as the underlying chemical and transport processes are individually complex and a thorough understanding of the coupled effects of these phenomena remains elusive. However, deeper insight into the mechanisms by which turbulent transport and reaction dynamics influence each other is essential for the future design of efficient systems.

In considering the evolution of the concentration of passive scalars in an incompressible flow, a one-way coupling between the fluid momentum and the scalars is typically assumed in writing the transport equations. In particular, if considering the concentrations of two scalars involved in a binary reaction system, the concentrations being denoted $C_1$ and $C_2$, in an imposed solenoidal velocity field, $U$, the governing equations can be formulated as

\begin{equation}
    \frac{\partial C_i}{\partial t} + \nabla \cdot \left( UC_i \right) = D_m \nabla^2 C_i - AC_iC_{j\neq i},
    \label{eqn:DNS_full}
\end{equation}

\noindent where $A$ is a reaction coefficient and $D_m$ is a scalar diffusivity which is assumed to be the same for both indexed scalars in this work. This equation is valid in the dilute limit, where the heat release from the reaction is so small as to affect neither the density nor the reaction kinetics. \cite{arvin_1997}

While this form of the equation incorporates all effects in the system, in practice, the Reynolds-averaged version of equation \ref{eqn:DNS_full}, written as 

\begin{equation}
    \frac{\partial \overline{C_i}}{\partial t} + \nabla \cdot \left( \overline{U}_{\,}\overline{C}_i \right) + \nabla \cdot \left( \overline{U'C_i'} \right) = D_m \nabla^2 \overline{C_i} - A\overline{C}_i\overline{C}_{j\neq i} - A\overline{C'_iC'_{j\neq i}},
    \label{eqn:RANS_full}
\end{equation}

\noindent has more utility due to its lower computational cost. In Eqn. \ref{eqn:RANS_full} primed terms denote fluctuations about the mean quantities, which are denoted by over-bars. In general, for a stochastic system, this averaging is done over multiple ensembles of flow realizations, and the corresponding set of partial differential equations for the velocity field are the Reynolds-averaged Navier Stokes (RANS) equations. Finding models for the two types of unclosed terms in Eqn. \ref{eqn:RANS_full}, $\overline{U'C_i'}$ and $\overline{C'_iC'_{j\neq i}}$, is an important problem in predictive science and for applications involving combustion \cite{peters_2000}, atmospheric and marine pollution \cite{seinfeld_2016}, electrolyte solutions in electrochemical applications \cite{Rosales_2017}, and other fields. In this work, such a model will be derived using a weakly-nonlinear extension of dispersion analysis.

\subsection{Taylor Dispersion}

For non-reacting scalar contaminants, many models exist for tracking the dispersion of passive scalars. The most common analytic model invoked for the unclosed transport term in the non-reactive equations, which is also used for some reactive cases, follows the generalized gradient diffusion hypothesis, for which the unresolved flux of a single passive scalar is of the form

\begin{equation}
    \overline{u_i'C_1'} = -D_{eff}\frac{\partial \overline{C_1}}{\partial x_i}, 
\end{equation}

\noindent where $D_{eff}$ is a generic effective eddy diffusivity that aggregates turbulent transport effects. When there is strong scale separation between the mean fields and the underlying fluctuating velocity fields, this local model is exact, with the caveat that $D_{eff}$ may be a non-isotropic tensor when the underlying fluctuating flow is not statistically isotropic. \cite{kraichnan_1987, mani_2021} 

In the framework of this model form, there is a rich body of literature on analytic models that find this effective diffusivity in canonical flows. In this study we denote the effective diffusivity in the absence of reaction as $D^0$, and refer to it as the "standard" eddy diffusivity. Correspondingly, we denote $D_{eff} = D^0$ as referring to the standard gradient diffusion model, which is agnostic to the reactivity of scalars.

There are many ways to measure this defining parameter. In particular, for the case of parallel flow in a pipe, \cite{taylor_1953} shows that a passive scalar experiences enhanced diffusive transport in the longitudinal direction. We can define a P\'eclet number as $Pe \sim uL/D_m$ such that, at high values of the P\'eclet number, $D^0$ dominates the standard molecular diffusivity, $D_m$. In parallel flows, it is generally shown that this quantity scales as

\begin{equation}
    \frac{D_{0}}{D_m} \sim \frac{u^2L^2}{D_m^2},
\end{equation}

\noindent where $u$ is the characteristic convective speed, and $L$ is some characteristic spanwise length scale for the flow. This idea can be expanded to applications outside of pipe flow, and other studies have expanded the range of problems and fields to ones that Taylor's original analysis does not address. \cite{taylor_1954,aris_1956, lighthill_1966, shapiro_1986} 

\subsection{Other Scalar Transport Models}

Thus far, we have only looked at non-reactive models. However, many tracers of interest are not passive; to reach a multi-component framework, the first step is the simplest possible reaction involving two species, where the concentration of the second species is orders of magnitude larger than the first everywhere in the domain. In some applications, this allows an assumption that the second species is maintained in the domain at a constant concentration and unaffected by the reaction dynamics, leading to a first-order linear reaction term for the first species. In the presence of such a first-order linear reaction, \cite{peters_2000} used work by \cite{corrsin_1961} to propose a model that incorporates the chemistry effects for the unclosed scalar flux term. It can be written as

\begin{equation}
    \overline{u_i'C_1'} = -\frac{D^0}{1+A\overline{C_2}\tau_{mix}}\frac{\partial \overline{C_1}}{\partial x_i}.
    \label{eqn:corrsin_flux}
\end{equation}

In this model, $\tau_{mix}$ is a characteristic time scale for mixing by the flow. In the denominator, one can form a Damk\"ohler number, relating the timescale of the reaction to the turbulence timescale, and can therefore write the flux as the gradient times an effective diffusivity that scales as 

\begin{equation}
\label{eqn:peters}
    D_{eff} \sim \frac{D_{0}}{1+Da},
\end{equation}

\noindent where $D^0$ is the aforementioned "standard" turbulent diffusivity. When the reaction rate is sufficiently slow or when examining the non-reacting limit where $Da$ vanishes, this recovers the gradient diffusion model exactly. While $D^0$ still needs to be determined from other means, the advantage of Eqn. \ref{eqn:peters} is that it analytically captures the impact of reaction on $D_{eff}$. 

Equation \ref{eqn:peters} contributes critical understanding by explicitly revealing the dependence of effective diffusivity (or eddy diffusivity) on the reaction kinetics and, specifically, the reaction coefficient. Although transport and kinetics are controlled by independent terms at the microscopic continuum level, when behavior at a macroscopic level, governed by ensemble-averaged equations, is sought, kinetics influence transport. 

In this study, we develop a model in the Reynolds-averaged context that can provide algebraic closures to the scalar evolution equations for a binary reactant setup. We do this using the spirit of Taylor's dispersion analysis. By considering the case of a parallel flow as an analytical prototype, we extend Taylor's analysis to discover model forms for dispersion of scalars undergoing binary reactions. Given the nonlinearity of this system, the closure operators are expected to be dependent on the transported quantity itself. We derive these dependencies through a weakly nonlinear extension of Taylor's original analysis. We then hypothesize that reactive mixing by turbulent flows should involve model forms with similar nonlinearities, but with different model coefficients. This is analogous to the situation of nonreactive scalars, where mixing by both classes of flows can be reasonably captured by gradient diffusion models. However, the diffusivity coefficients differ based on the underlying flow topology, with $D^0/D_M$ scaling as $Pe^2$ in parallel flows and as $Pe^1$ in turbulent flows.  

Our weakly nonlinear model offers improvements to perturbation expansions based on directly linearizing the equations performed by works like \cite{battiato_2011}. It also recovers the scaling relationship between diffusivity and Damk\"ohler number derived in \cite{elperin_2014} and \cite{elperin_2017}, although those works use nonlinear first-order reactions. It also indicates that the effective eddy diffusivity is linked to reaction, as has been shown in experiments. \cite{watanabe_2014}  This relationship shows that the impacts of reaction chemistry and turbulent mixing are fundamentally linked. 

In what follows, we will first develop a model problem based on laminar flows that can capture the relevant multi-physics at play for a linear reaction setup and work to find a solution for the single-scale and the multi-scale problems. This insight will guide generalization to a binary reaction problem, which we can fully expand to general turbulent flows. The outcome of our analysis is a unifying reduced-order model (ROM) in the RANS context that captures the analytic impact of chemical kinetics on transport and reaction closures while leveraging existing literature on non-reactive scalar transport modeling.

\section{Model Problem}

\begin{figure}
    \centering
    \includegraphics[width=0.43\textwidth,angle=90]{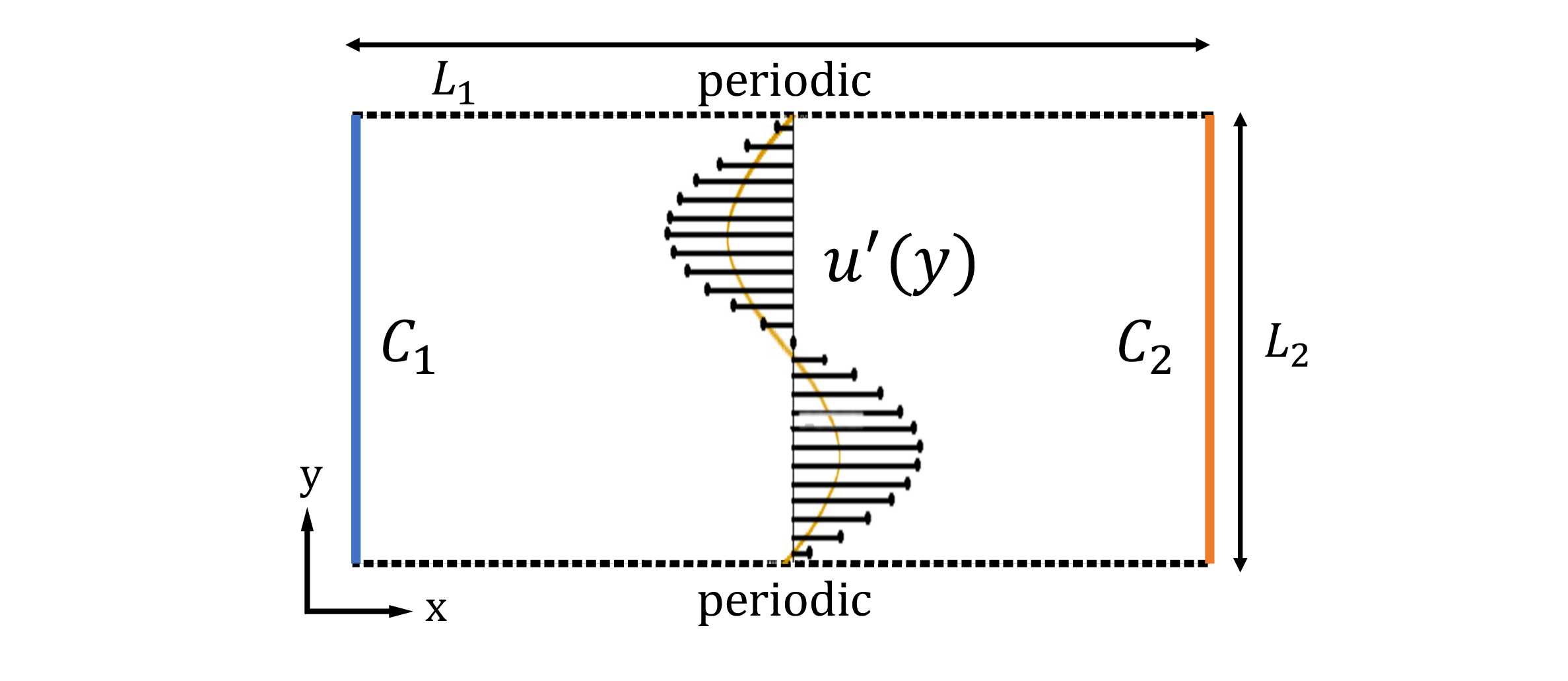}
    \caption{A schematic of the model problem, illustrating dimensions and boundary conditions. The flow is depicted in black as $u'(y)$, with a representation of a reaction front given at the mid-plane. Axial will refer to the $x$-direction, while spanwise will refer to the $y$-direction}
    \label{fig:cartoon}
\end{figure}{}

In order to develop insights into a model form that captures closure terms in a binary mixture, we propose the following illustrative setup that can be treated semi-analytically. Consider a flow field that is parallel, steady, and two-dimensional, contained in a domain that is an elongated box. We place periodic boundary conditions on the two opposing elongated sides, as seen in Fig. \ref{fig:cartoon}. By elongated, we mean that $L_1 >> L_2$.

In such a two-dimensional domain, we define an averaging operator as 

\begin{equation}
    \overline{f(x)} = \frac{1}{L_2} \int_{0}^{L_2} f(x,y) dy, 
\end{equation}

\noindent which differs from the ensemble averaged used earlier. In this context, let us use a periodic structure for the flow-field which imposes no mean velocity. Namely, $u(x,y) = \overline{u} + u' = U_0 sin(ky)$, where $k = 2\pi/L_2$. At $x = -L1/2$ and $x = L_1/2$, the two domain boundaries, we prescribe Dirichlet conditions for the scalar concentrations.

In addition, we consider a reaction of the form $C_1 + C_2 \rightarrow C_3$, which is considered irreversible and fully activated, but follows a finite rate law of mass action. While simpler than a fully three-dimensional turbulent flow with reacting scalars, this model problem captures the essential competing physics of mixing, transport, and reactions that are present in the more realistic case. However, we can now formulate solutions in this sandbox context and propose closures for each sub-problem that each highlight different aspects of the overall ROM.

\subsection{Linear Reaction, Single-Scale Flow}\label{sec:lr-ssf}

Let us first examine this problem in the context of a linear reaction. In this setup, the concentration of one of the scalars, $C_2$, is held constant in the entire domain at unity and does not evolve, while the maximum value of $C_1$, at the left side of the domain, is held to be at least an order of magnitude less than that of $C_2$. As $C_2$ is a constant, we will denote the reaction rate as $A_L=AC_2$, denoting the product of a standard binary species reaction coefficient times the concentration of the constant species. The fluctuations for $C_1$ are governed by a transport equation as

\begin{equation}
    \frac{\partial C_1'}{\partial t} + \frac{\partial (u'C_1')'}{\partial x_1}  + u'(y) \frac{\partial \overline{C_1}}{\partial x_1} = D_m \frac{\partial^2 C_1'}{\partial x_i \partial x_i} -  A_L C_1', 
    \label{eqn:linear_fluc}
\end{equation}

\noindent where $C_1 = \overline{C_1} + C_1'$ and so on. This equation can be derived by finding the classical transport equation for $C_1$ and subtracting from it the evolution equation for $\overline{C_1}$. Note that we have implicitly assumed here that the molecular diffusivity, $D_m$, is the same for all species, is constant, and is isotropic.

This equation is similar to the equation governing evolution of fluctuating fields in the work of \cite{taylor_1953}, with the primary distinction being the reaction term on the right hand side. While Taylor was not explicitly considering this type of periodic parallel flow in his work, the steps he follows allow us to also simplify this equation to arrive at a more analytically approachable differential equation. 

First, we note that we are considering a domain where $L_1 >> L_2$, from which we can conclude that diffusion in the axial direction, $x$, can be neglected when compared to diffusion in the spanwise direction, $y$. This can be shown by performing a scaling analysis and seeing the associated term tends to zero as the relevant P\'eclet number tends to infinity. This allows us to neglect axial diffusion in Eqn. \ref{eqn:linear_fluc}. 

In addition, when $L_1 >> L_2$, the time scale of mixing in the axial direction is much larger than the time scale of mixing in the spanwise direction. Therefore, one can decompose the concentration field as $C_1(x,y) = \overline{C_1}(x) + C_1'(x,y)$, allowing for evolution of the mean field in the axial direction but considering the underlying field to be "well mixed" in the spanwise direction. In particular, the presence of fast mixing in the spanwise direction implies that $|C_1'| <<  |\Delta \overline{C_1}|$, where $|\Delta \overline{C_1}|$ is a characteristic change in the averaged concentration in the axial direction.

This final conclusion allows us to see that the second term on the left-hand side of Eqn. \ref{eqn:linear_fluc} must be much less significant than the advection of the mean scalar concentration, represented by the third term on the left-hand side. Therefore, for a leading-order approximation, we can neglect the former, double primed term.

All of these physical assumptions imply that the flow is in an advection-driven limit. The strong mixing in the spanwise direction that such a flow implies also allows us to invoke a quasi-steady approximation for the evolution of the fluctuating concentration. In particular, since we are most interested in the long-time response, not the initial transient one, this assumption does not unreasonably limit our analysis. 

If the previous simplifications are applied to Eqn. \ref{eqn:linear_fluc}, the dominant remaining terms give us an approximate evolution equation that can be written as

\begin{equation}
   u'(y) \frac{\partial \overline{C_1}}{\partial x} = D_m \frac{\partial^2 C_1'}{\partial y^2} -  A_L C_1'.
\end{equation} 

\begin{figure}
    \centering
    \includegraphics[width=0.75\textwidth]{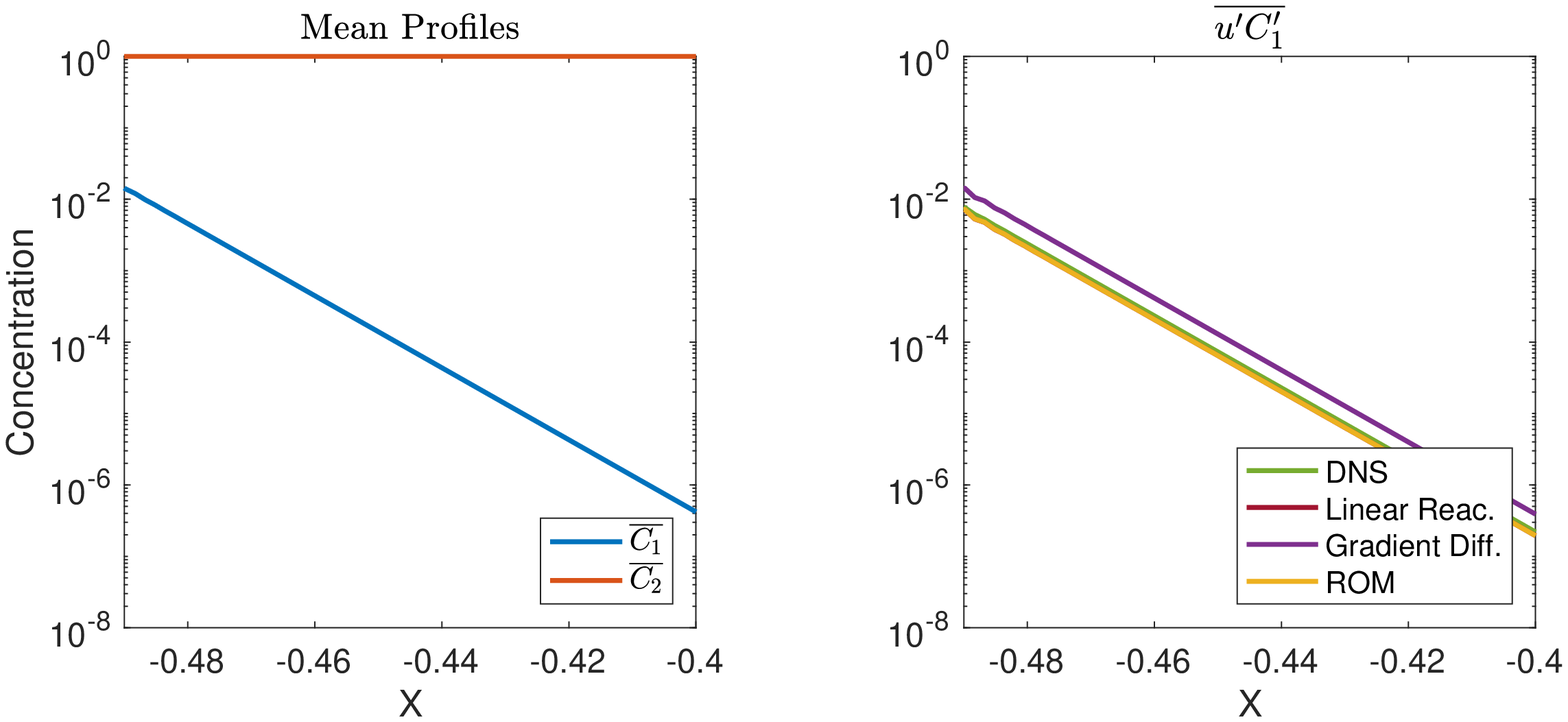}
    \caption{Predictions of closures for the linear reaction problem subject to a turbulent flow, $X = x/L_1$. The Linear Reaction model of Eqn. \ref{eqn:corrsin_flux} is identical to the ROM represented by Eqn. \ref{eqn:ROM_linear_flux}, and the former is overlaid by the latter in the plot.
}
    \label{fig:lr}
\end{figure}{}

This gives us the dispersion of the fluctuating quantity as driven by the flow. A further ansatz we make here is that the form of the fluctuating components follows the velocity, as $C_1(x,y) = \overline{C_1}(x) + f_1(x)sin(ky)$. That is, the fluctuating field has some magnitude determined purely by axial position. This is motivated by examining solely the leading-order term in the harmonic expansion of the concentration fluctuations in the $y$-direction. Substituting this into the fluctuation equation yields 

\begin{equation}
    D_m k^2 f_1(x) + A_L f_1(x) = -U_{0} \frac{\partial \overline{C_1}}{\partial x}.
\end{equation}

Solving this equation for our unknown axial concentration magnitude yields

\begin{equation}
    f_i(x) = -\frac{U_0}{k^2D_m+A_L}\frac{\partial \overline{C_i}}{\partial x}.
\end{equation}{}

The only unclosed term in Eqn. \ref{eqn:linear_fluc} is the scalar flux, and it can now be solved as

\begin{equation}
    \overline{u'C_1'} = \overline{U_0f_1(x)sin^2(ky)} = - D_{eff} \frac{\partial \overline{C_1}}{\partial x}.
    \label{eqn:ROM_linear_flux}
\end{equation}

Here, $D_{eff}$ is given by 

\begin{equation}
    D_{eff} = \frac{D_{0}}{1+A_L\tau_{mix}},
    \label{eqn:deff_lrss}
\end{equation}

\noindent where the standard eddy diffusivity, $D^0$, is determined from non-reactive Taylor-type analysis. For this flow, it is given by 

\begin{equation}
    D^0 = \frac{U_0^2}{2D_mk^2}.
\end{equation}

In addition, $\tau_{mix}$ is a characteristic timescale for mixing given by 

\begin{equation}
    \tau_{mix} = \frac{1}{D_mk^2} = \frac{D_{0}}{u_{rms}^2},
    \label{eqn:deff_tau}
\end{equation}

\noindent where $u_{rms}$ is the root-mean-squared velocity of the underlying flow, equal to $U_0/\sqrt{2}$ in this derivation. For this section, we will refer to Eqn. \ref{eqn:corrsin_flux} as the Linear Reaction model and Eqn. \ref{eqn:deff_lrss} as a special case of our Reduced-Order Model (ROM) for this first-order reaction context.

Interestingly, for this problem, the ROM is identical to Eqn. \ref{eqn:corrsin_flux}, even though it is derived in a laminar context, while the latter was created in the framework of turbulent flows. It again has an effective Damk\"ohler number in the denominator, and shows that effective diffusivity decreases as the reaction rate increases relative to the flow timescale. This lends confidence that the choice of model problem  should not hinder the relevance of the conclusions here in application to turbulence. 

In particular, the model form is an explicit function of $A_L$, with the other two quantities, $D^0$ and $u_{rms}$, defined purely in terms of flow parameters. This means that they can be derived independent of the reaction kinetics, in non-reacting flow, and applied directly to this model for reacting flows.

To test this model, we next examine its predictions against 2D DNS data as well as the gradient diffusion model that ignores impact of reactions on the advective closure. For this purpose, we consider a setting with $Pe \equiv (U_0 L_2)/(2 \pi D_m)=100$, and $Da \equiv A_L \tau_{mix} = 1$. For the axial boundary conditions, we used Dirichlet conditions $x = -L_1/2$ and $x=L_1/s$ by respectively enforcing $C_1$ to be equal to $0.1$ and $0$.    

Figure \ref{fig:lr} presents quantitatively a comparison between different results. For this specific setting, the mean concentration assumes an exponential profile, as seen in the figure. It can be seen that ROM produces results much closure to the DNS than the gradient diffusion model. The gap between DNS and gradient diffusion model is solely due to over prediction of $D_{eff}$ by the model. For settings at higher $Da$ the correction offered by ROM becomes even more crucial as this gap will be even larger.



\subsection{Linear Reaction, Multi-Scale Flow}\label{sec:lr-msf}

The previous section only considered a velocity field with a single relevant length scale. However, a realistic flow has a spectrum of relevant scales and the model problem setup allows us to consider a straightforward superposition of multiple velocity fields. The total velocity now can be represented by $u(x,y) = \sum U_n sin(nky)$, where $k = 2\pi/L_2$ and $n$ refers to each scale as corresponding to an integer multiple of the fundamental scale set by $k$. Following the same procedure as the previous section, the unresolved scalar flux can be expressed as

\begin{equation}
    \overline{u'C_1'} = - \sum_{n=1}^\infty \frac{U_n^2/2}{n^2k^2D_m+A_L} \frac{\partial \overline{C_1}}{\partial x}.
    \label{eqn:multi_scale}
\end{equation}

In this context, the dependence of $D_{eff}$ on $A_L$ is more complex. Nevertheless, we still seek to express the dependence of $D_{eff}$ on $A$ in terms of a simple expression that involves low order statistics, specifically $D^0$ and $u_{rms}$. In order to do so, we first examine the expression above in the asymptotic limits of large and small $A_L$. In the limit of $A_L$ equaling zero, the above expression results in $D_{eff}$ equal to the $D^0$ of the multi-scale flow field, recovering the non-reactive solution from pure Taylor-type analysis. In the limit of large $A_L$, however, we obtain that $D_{eff} = u_{rms}^2/A_L$. 

\begin{figure}
    \centering
    \includegraphics[width=.6\textwidth]{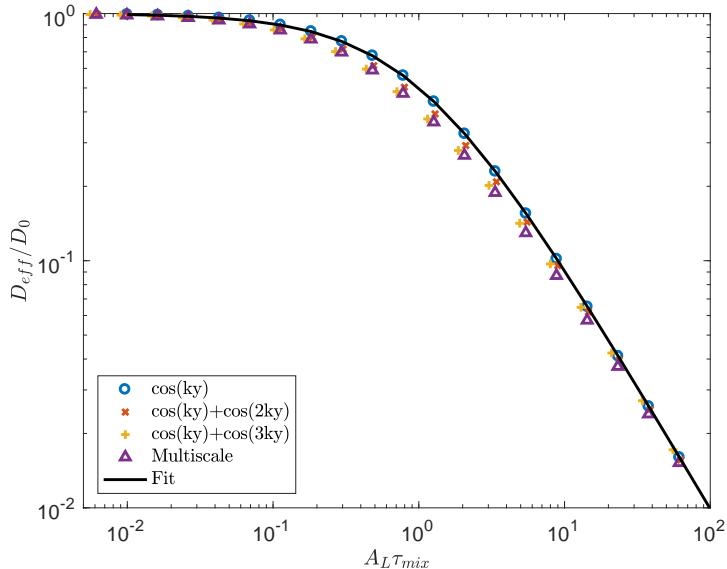}
    \caption{Effective diffusivity as a function of $Da$. Each marker refers to a different velocity field, with the "Multiscale" field given by Eqn. \ref{eqn:multi_scale_velocity}. The solid line represents Eqn. \ref{eqn:multi}.}
    \label{fig:lrms_deff}
\end{figure}{}

Interestingly, examining Eqn. \ref{eqn:deff_lrss} and updating values for $D^0$ and $\tau_{mix}$ satisfies both limits exactly. 

\begin{equation}
    D_{eff} = \frac{ D^0 }{1 + A_L \tau_{mix}},
    \label{eqn:multi}
\end{equation}

\noindent where now 

\begin{equation}
    D^0 = \sum_{n=1}^\infty \frac{U_n^2/2}{n^2k^2D_m}
\end{equation}

\noindent and $\tau_{mix}$, the mixing time, is $D^0/u_{rms}^2$, where 

\begin{equation}
    u^2_{rms} = \sum_{n=1}^\infty U_n^2.
\end{equation}

We reiterate that both $D^0$ and $u_{rms}$, and therefore $\tau_{mix}$, are fundamental measures of the flow that are independent of the presence of the reaction, and there is no $A_L$ dependence in these quantities. 

One can examine this model by comparing the $D_{eff}$ it provides against those from Eqn. \ref{eqn:multi_scale}, as shown in Fig.\ref{fig:lrms_deff}. Among the various velocity profiles tested, we considered the the following velocity profile, denoted as ``Multi-scale" in Figure~\ref{fig:lrms_deff}, expressed as 

\begin{equation}
u'(y)=\sum_{n=1}^{10} n^{-5/6}cos(nky),
\label{eqn:multi_scale_velocity}
\end{equation}

\noindent to mimic the multi-scale effects found in turbulence. The specific power, $-5/6$, is chosen such that the eddy diffusivity at each wavenumber matches that of Kolmogorov turbulence. 

In Fig. \ref{fig:lrms_deff}, the simple model presented in Eqn. \ref{eqn:multi} is plotted across values of the reaction rate. We see that it captures the precise behavior of $D_{eff}$ in the limits of small and large $A_L$ across a variety of imposed velocity fields. Additionally, the model still fits the data reasonably in the intermediate regimes where $A_L\tau_{mix} \sim O(1)$.  

\subsection{Binary Reaction}

Having some confidence with the linear problem, we now turn back to examining Eqn. \ref{eqn:RANS_full} in its entirety, and note that the complete problem involves a full binary system, where both $C_1$ and $C_2$ are free to evolve.

Before proceeding with derivation of a closure model, let us first test whether we can naively modify Eqn. \ref{eqn:corrsin_flux} to perform reasonably in the case of binary reactions. Specifically, one may interpret that for evolution of each species, $C_i$, the other species might be assumed "locally constant." This allows interpreting an effective $A_L$ for each species as $A_L=A\overline{C}_{j\ne i}$. With this simple modification, the previous model form can be extended to a multi-scale model as 

\begin{equation}
    \overline{u'C_i'}  = -\frac{D}{1+A\tau\overline{C_j}}\frac{\partial \overline{C_i}}{\partial x}.
    \label{eqn:linear-extended}
\end{equation}{}

\noindent where $A_L$ is now replaced with the $A\overline{C_2}$ and  $A\overline{C_1}$ for mean advective fluxes associated with $C_1$ and $C_2$, respectively.  

We next examine the performance of this model form by comparing its prediction against direct numerical simulation. In this case we consider a single-scale velocity profile where $u(y) = U_0 sin(ky)$, but where both reactants are now free to vary. We choose $Pe \equiv (U_0 L_2)/(2 \pi D_m) = 100$ and $Da \equiv A \tau_{mix} C_{ref} = 100$, where $C_{ref}$ is concentration imposed by Dirichlet boundary conditions on each scalar. More specifically, the boundary condition on $C_1 = C_{ref}$ is imposed at $x = -L_1/2$ and $C_2 = C_{ref}$ is imposed at $x = L_1/2$. The other boundary condition for each scalar is a homogeneous Dirichlet condition.

\begin{figure}
    \centering
    \includegraphics[width=0.75\textwidth]{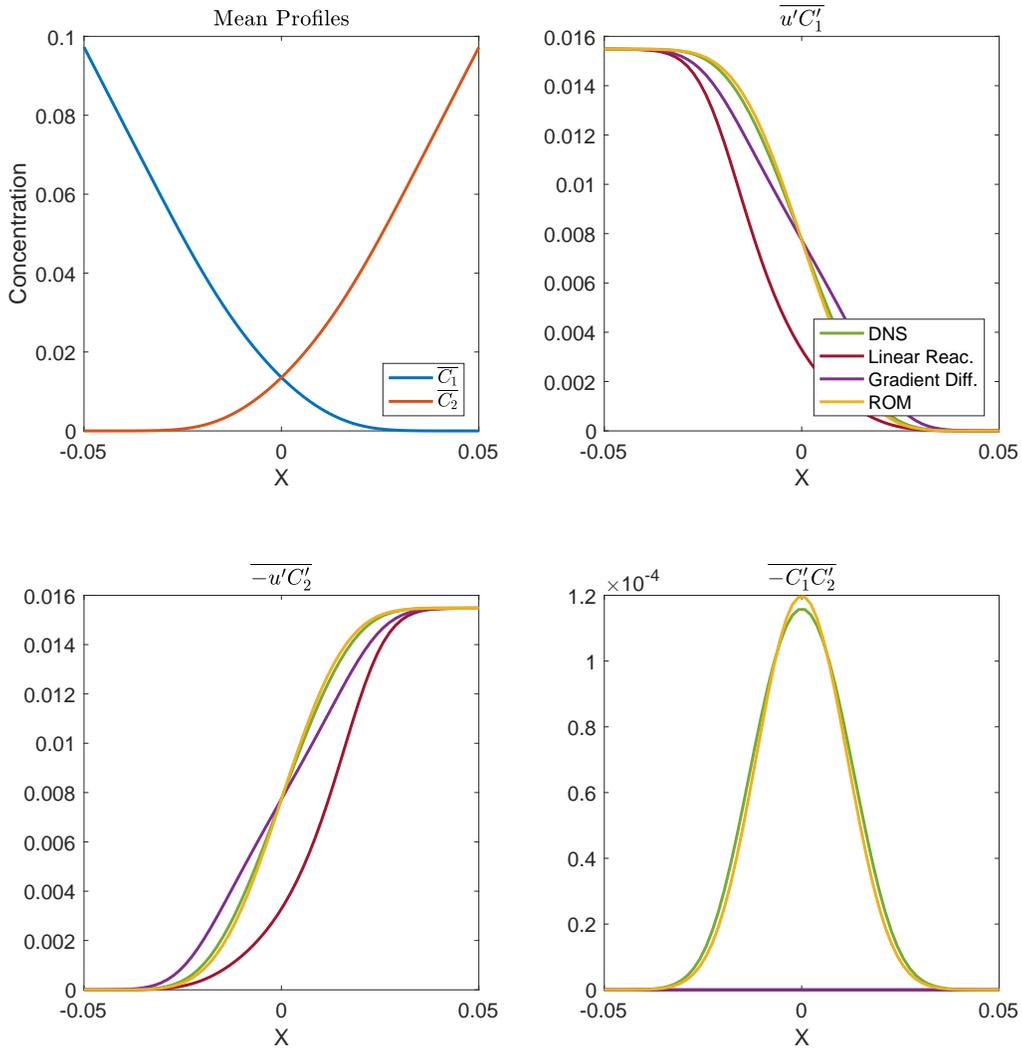} 
    \caption{Predictions of closures for the binary reaction problem subject to a laminar flow, $X = x/L_1$.}
    \label{fig:dns}
\end{figure}

In Fig. \ref{fig:dns}, we see that the Linear Reaction model given by Equation \ref{eqn:linear-extended} incurs great errors near the middle of the domain, where $C_1$ and $C_2$ exist in near-stoichiometric ratios. Surprisingly, the standard Gradient Diffusion model, which carries no chemistry information, now outperforms this more physics-based model. 

{}

Clearly, the Linear Reaction model is not sufficient to describe this system, so let us return to Eqn. \ref{eqn:RANS_full} and examine the full system of equations again. In this framework, the governing equations for the fluctuations control the evolution of the unclosed terms of interest. For $C_1$, this looks like  

\begin{equation}
    \frac{\partial C_1'}{\partial t} + \frac{\partial (u'C_1')'}{\partial x_1}  + u'(y) \frac{\partial \overline{C_1}}{\partial x_1} = D_m \frac{\partial^2 C_1'}{\partial x_i \partial x_i} \\
    - A\overline{C_1}C_2' -  A\overline{C_2}C_1' \\
    + A(C_1'C_2')'.
    \label{eqn:binary_fluc}
\end{equation}

Again following the work of \cite{taylor_1953}, we can add one more assumption for the binary system that was not relevant for the linear case, which is that the fluctuations of the product of fluctuating quantities will be negligible when compared to the mean field quantities.

If those simplifications are applied to Eqn. \ref{eqn:binary_fluc} for $C_1$, it can be reformulated as

\begin{equation}
   u'(y) \frac{\partial \overline{C_1}}{\partial x} = D_m \frac{\partial^2 C_1'}{\partial y^2}
    - A\overline{C_1}C_2' -  A\overline{C_2}C_1',
\end{equation}

\noindent and an equivalent equation can be derived for $C_2$.

We again make the ansatz that $C_i(x,y) = \overline{C_i}(x) + f_i(x)sin(ky)$, following the velocity. Substituting this into the fluctuation equations yields the following system of two equations:

\begin{equation}
    D_m k^2 f_1(x) + A\overline{C_1} f_2(x) +  A\overline{C_2} f_1(x) = -U_{0} \frac{\partial \overline{C_1}}{\partial x}
\end{equation}

and 

\begin{equation}
    D_m k^2 f_2(x) + A\overline{C_2} f_1(x) +  A\overline{C_1} f_2(x) = -U_{0} \frac{\partial \overline{C_2}}{\partial x}.
\end{equation}

This is a coupled, but linear, set of equations for $f_1$ and $f_2$ that can be solved to find that

\begin{equation}
    f_1(x) = \frac{-U_{0}(k^2D_m + A\overline{C_1})}{ k^2D_m(k^2D_m+A\overline{C_1}+A\overline{C_2})}\frac{\partial \overline{C_1}}{\partial x} + \frac{U_{0}(A\overline{C_1})}{ k^2D_m(k^2D_m+A\overline{C_1}+A\overline{C_2})}\frac{\partial \overline{C_2}}{\partial x}
\end{equation}

and

\begin{equation}
    f_2(x) = \frac{-U_{0}(k^2D_m + A\overline{C_2})}{ k^2D_m(k^2D_m+A\overline{C_1}+A\overline{C_2})}\frac{\partial \overline{C_2}}{\partial x} + \frac{U_{0}(A\overline{C_2})}{ k^2D_m(k^2D_m+A\overline{C_1}+A\overline{C_2})}\frac{\partial \overline{C_1}}{\partial x} .
\end{equation}

We can also write these terms, the result of using a weakly-nonlinear extension of dispersion analysis, as the superposition of two gradients multiplied by a prefactor representing an effective diffusivity. The equations for the scalar flux now look like 

\begin{equation}
\begin{bmatrix}
\overline{u' C_1'} \\
\overline{u' C_2'}
\end{bmatrix}
=-
\begin{bmatrix}
D_{11} & D_{12} \\
D_{21} & D_{22}
\end{bmatrix}
\frac{\partial}{\partial x}
\begin{bmatrix}
\overline{C_1} \\
\overline{C_2}
\end{bmatrix}
,
\label{eqn:ROM_transport}
\end{equation}{}   

\noindent where $D_{kl}$ represents a diffusivity associated with the flux of the $k$-th species due to gradient in the species. The square matrix formed by these coefficients is also positive definite so long as the mean scalar fields maintain their positivity. The coefficients are written as, 
    
\begin{align}
\begin{split}
&D_{11} = \frac{D^0(1 + A\overline{C_1}\tau_{mix})}{1+A\tau_{mix}\overline{C_1}+A\tau_{mix}\overline{C_2}}  \;\;\;\;\;\;     D_{12} = -\frac{D^0A\tau_{mix}\overline{C_1}}{1+A\tau_{mix}\overline{C_1}+A\tau_{mix}\overline{C_2}} \\
&D_{21} = -\frac{D^0A\tau_{mix}\overline{C_2}}{1+A\tau_{mix}\overline{C_1}+A\tau_{mix}\overline{C_2}}  \;\;\;\;\;\;     D_{22} = \frac{D^0(1 + A\overline{C_2}\tau_{mix})}{1+A\tau_{mix}\overline{C_1}+A\tau_{mix}\overline{C_2}}
\end{split}
\label{eqn:Ddefinitions}
\end{align}

\noindent using the notation from Section \ref{sec:lr-ssf}.

Perhaps most novel, however, is that determining the actual local form of the fluctuations means it is now possible to close not only the scalar transport term, but also the unclosed reaction terms in the standard RANS equations. This closure equation looks like 

\begin{equation}
    \overline{AC_1'C_2'} = \frac{A}{u_{rms}^2} \left( D_{11}\frac{\partial\overline{C_1}}{\partial x} + D_{12} \frac{\partial \overline{C_2}}{\partial x} \right) \left( D_{21}\frac{\partial\overline{C_1}}{\partial x} + D_{22} \frac{\partial \overline{C_2}}{\partial x} \right). 
    \label{eqn:ROM_scalar}
\end{equation}

 This provides our Reduced-Order Model (ROM) in its most complete form. It solves a reduced-degree-of-freedom system resulting from the RANS equations. The three algebraic components describe an entire model form, and require only prescriptions of parameters of the underlying flow, specifically $D^0$ and $u_{rms}$. It is important to note that this model, under the appropriate limits, captures the Gradient Diffusion model and the Linear Reaction as its special cases. The Gradient Diffusion model is realized in the limit of $A \rightarrow 0$. The Linear Reaction model, for finite $A_L$, can be realized in the limit of $\overline{C_2} \rightarrow \infty$ and $A \rightarrow 0$.

Comparing this model against the DNS data presented in Fig.\ref{fig:dns}, we see that the new model recovers from the errors of the Linear Reaction model, and outperforms the Gradient Diffusion model. 

The multi-scale analysis performed in Section \ref{sec:lr-msf} is omitted here as it would yield a similar summation form as in that section, but as seen in the linear regime, a model form from the single-scale case is still valid.

\section{Application to Turbulent Flows}

Now that we have built up the full ROM, we can turn to realistic turbulent flows. A great strength of the approach used in the previous section is that we can borrow the model form from the laminar analysis directly and use it in an analogous domain of homogeneous, isotropic turbulence. For this flow configuration, as a recapitulation, the ROM is given by Eqns. \ref{eqn:ROM_transport} through \ref{eqn:ROM_scalar}, with the only difference that flow parameters $D^0$ and $\tau_{mix}$ are now properties of a turbulent flow. As the ROM captures reaction-dependent effects explicitly, these parameters are agnostic to the presence of a reacting system and non-reactive flow measurements provide the values needed.

Specifically, \cite{mani_2021} explicates the process of finding these values for non-reactive turbulent flows using the Macroscopic Forcing Method (MFM). In particular, \cite{shirian_2020} used MFM to measure the true non-local form of the macroscopic eddy diffusivity, and additionally found values for $D^0$ in the purely local and isotropic limit applicable for this work.

\subsection{\textit{A priori} analysis}

\begin{figure}
\begin{subfigure}{.45\textwidth}
  \centering
  \includegraphics[width=\linewidth]{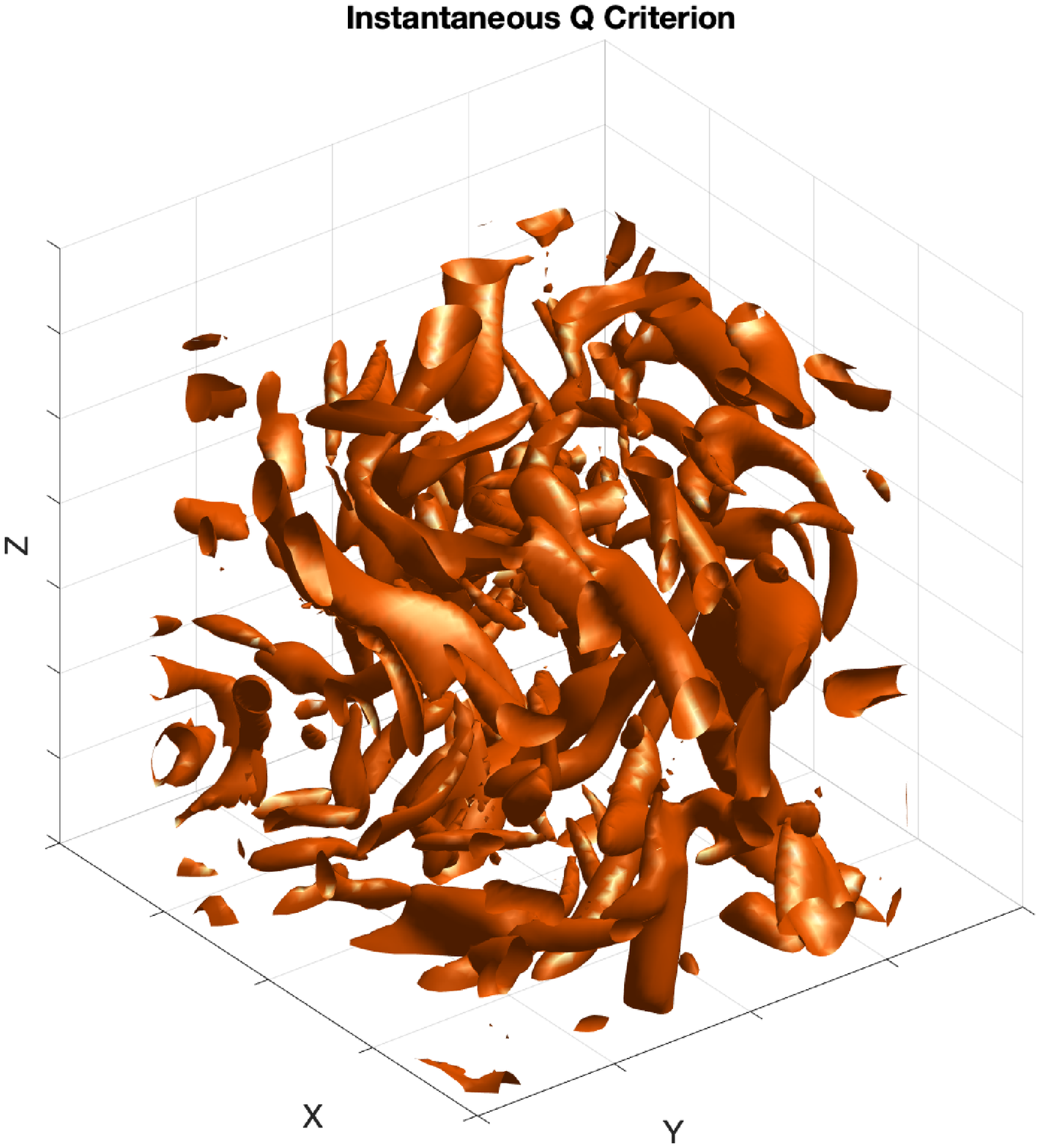}  
  \caption{ }
\end{subfigure}
\begin{subfigure}{.45\textwidth}
  \centering
  \includegraphics[width=\linewidth]{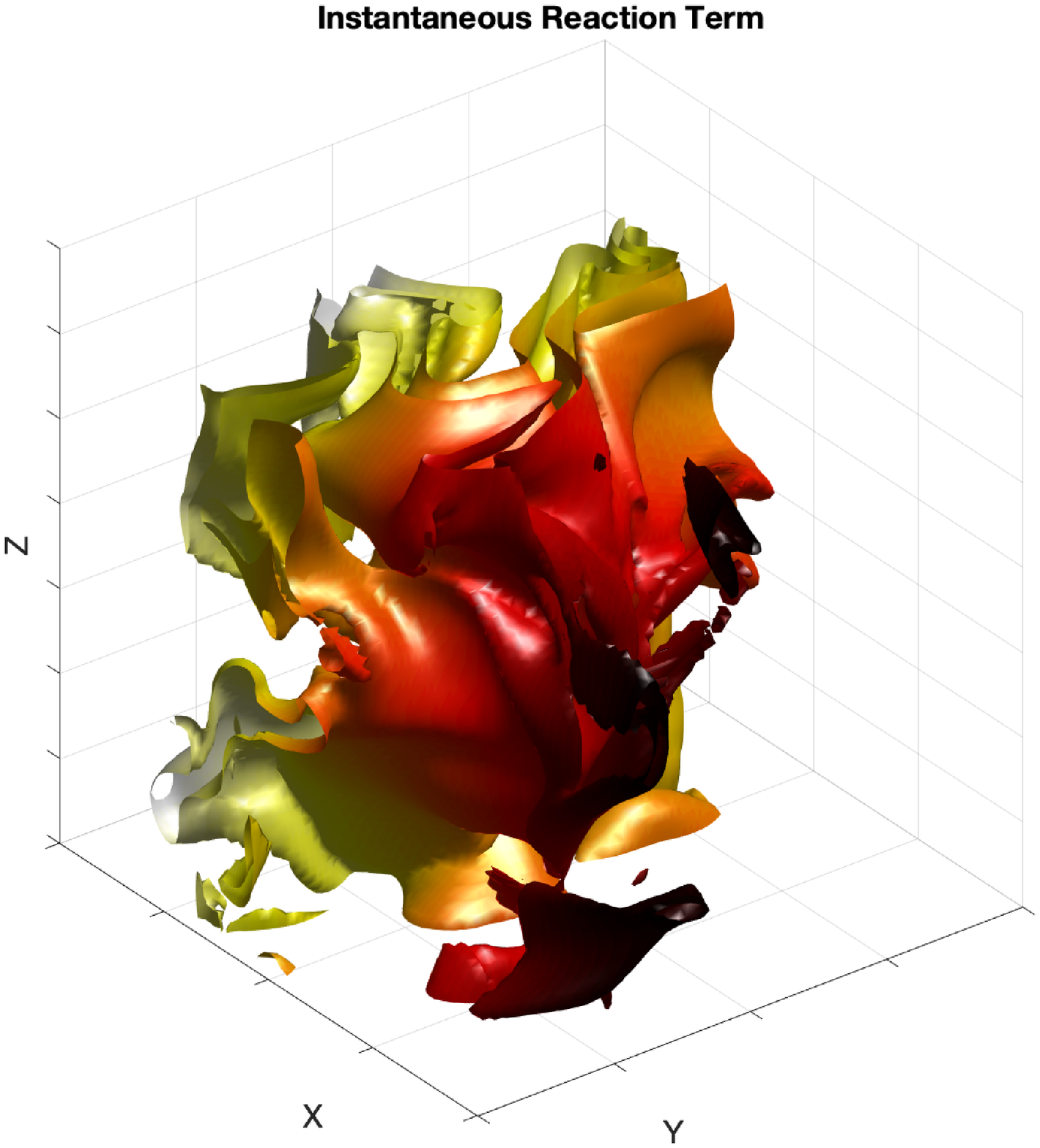}
  \caption{ }
\end{subfigure}
\caption{Instantaneous visualization of a binary reaction system subject to homogeneous, isotropic turbulence (a) An isosurface of the instantaneous $Q$ criterion that indicates eddy cores (b) An isosurface of the instantaneous reaction term, $A C_1 C_2$, colored by axial coordinate. In both figures, axes are equally scaled.}
\label{fig:turb_vis}
\end{figure}

In particular, we replace the steady, parallel flow examined in the previous section with three dimensional (3D) homogeneous isotropic turbulence (HIT) in an elongated domain to generalize the model problem. The code of \cite{pouransari_2016} in an incompressible mode was adapted for this work to simulate HIT in a 3D domain of size $(2 \pi)^3$. The resulting flow field was periodically extended in the $x$-direction to generate a computational domain for scalar transport with dimensions $20\pi\times2\pi\times2\pi$. The scalar transport equations are solved with boundary conditions identical to the 2D binary reaction problem. This provides a realistic reaction zone in the middle of the domain, far from the boundaries, as seen in Fig. \ref{fig:turb_vis}. 

For our flow simulations, we adopted parameters from \cite{shirian_2020} for a case of $Re_{\lambda} = 26$, which was obtained by setting the HIT box size of $L = 2 \pi$, the turbulent forcing parameter following the prescription of \cite{rosales_2005} equal to $0.2792$, and the kinematic viscosity to $\nu = 0.0263$.

This setup allows adoption of the values for non-reactive eddy diffusivity. For this specific case, \cite{shirian_2020} reported that $D^0 = 0.86u_{rms}^4/\epsilon = 0.96$ and $\tau_{mix} = 0.86u_{rms}^2/\epsilon = 1.02$, where $\epsilon = 0.79$ is the dissipation rate of turbulent kinetic energy and $u_{rms} = 0.97$ is the single-component root-mean-squared velocity. 

For scalar transport, we consider molecular diffusivity $D_m = 0.0263$, matching $\nu$, $C_{ref} = 1$ and a reaction rate $A = 100$. This leads to $Pe \equiv D^0/D_m = 37$ and $Da \equiv A \tau_{mix} C_{ref} = 102$.

Figure \ref{fig:turb_vis} shows imagery captured near the middle of the computational domain. Figure \ref{fig:turb_vis}a shows vigorous turbulent structures that mix the flow and the scalar fields. In response, reaction fronts are not planar; instead, one can observe highly stretched and distorted structures. Similar to the laminar problem, however, the ensemble-averaged fields will be smooth and one-dimensional.

\begin{figure}
    \centering
    \includegraphics[width=0.75\textwidth]{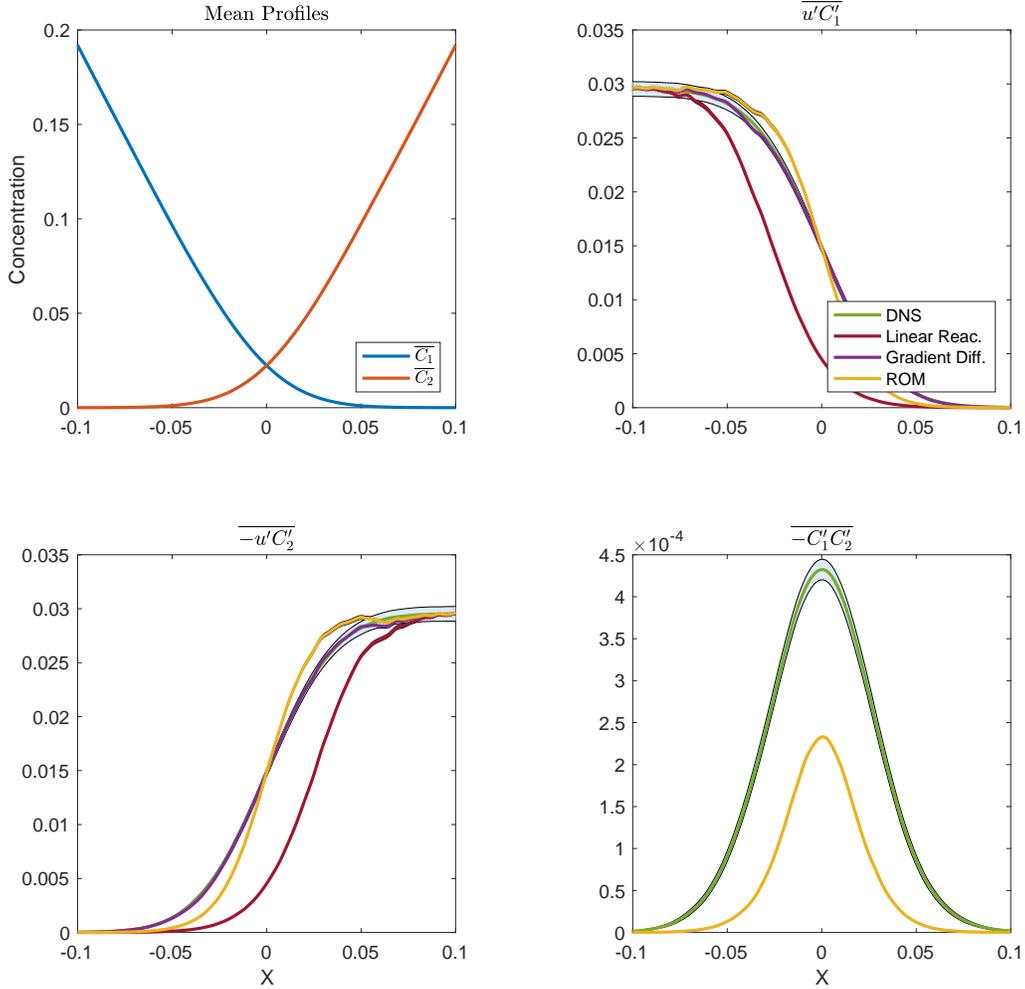}
    \caption{Predictions of closures for the binary reaction problem subject to a turbulent flow. $X = x/L_1$ and DNS results are presented with 95\% confidence intervals accounting for statistical fluctuations.}
    \label{fig:turb_full}
\end{figure}{}

In Fig.\ref{fig:turb_full}, the closure terms for the ensemble-averaged binary reaction problem are examined in the defined turbulent context. Similar to the laminar binary reaction problem, the full ROM recovers from the errors in capturing the true transport closures that are incurred by the Linear Reaction model of Eqn. \ref{eqn:linear-extended}. For these specific closure terms, the standard Gradient Diffusion model appears to match the DNS data more closely than the ROM. However, an overall assessment that considers both reaction and transport closures reveals the advantage of the ROM, as the Gradient Diffusion transport closure model offers no answers to the equally vital reaction closure question. Quantifying this advantage can be accomplished by performing \textit{a posteriori} analysis and comparing 1D model predictions against DNS data.

\subsection{\textit{A posteriori} analysis}

\begin{figure}
\begin{subfigure}{.494\textwidth}
  \centering
  \includegraphics[width=\linewidth]{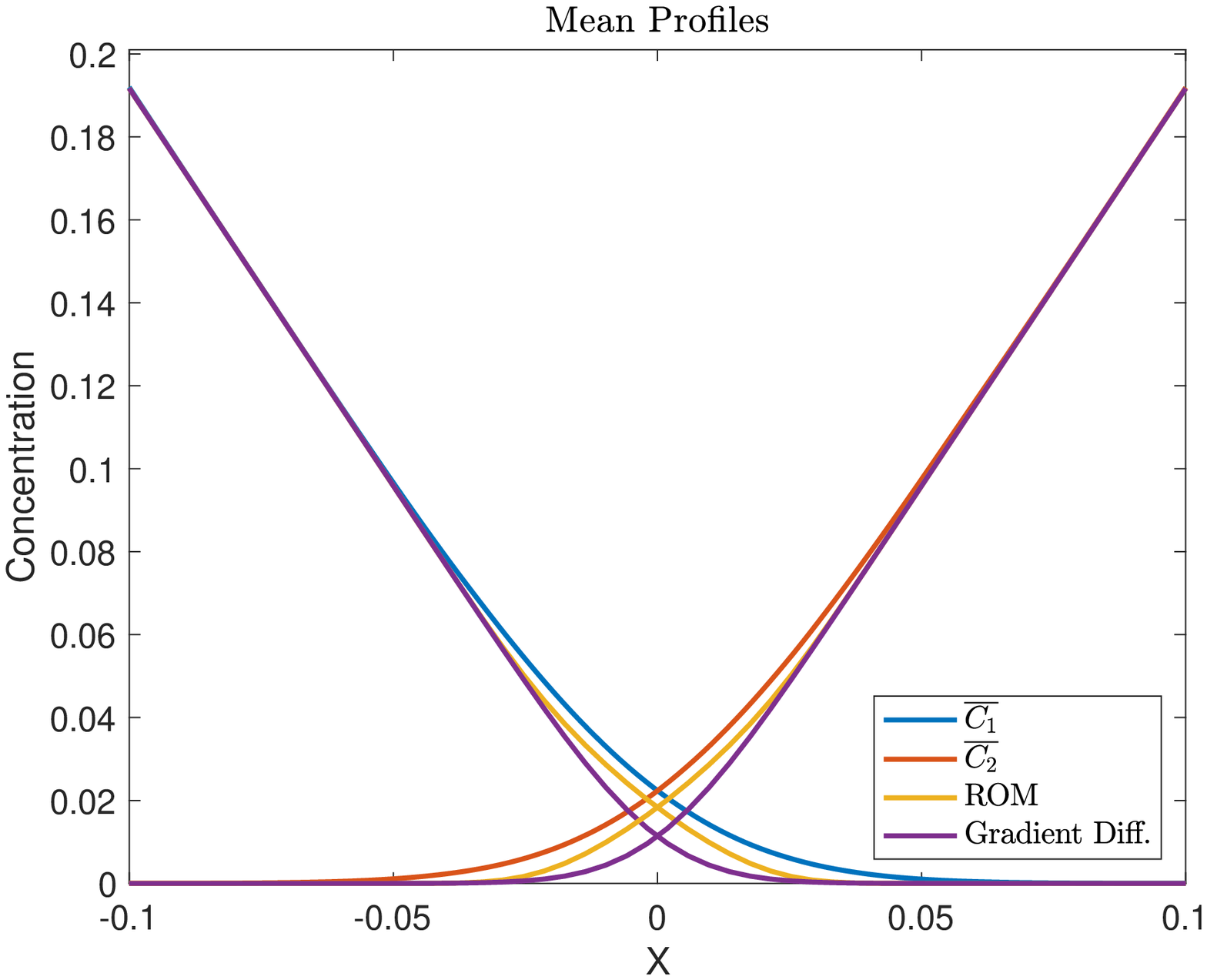}  
  \caption{ }
    \label{fig:RANS_full}
\end{subfigure}
\begin{subfigure}{.494\textwidth}
  \centering
  \includegraphics[width=\linewidth]{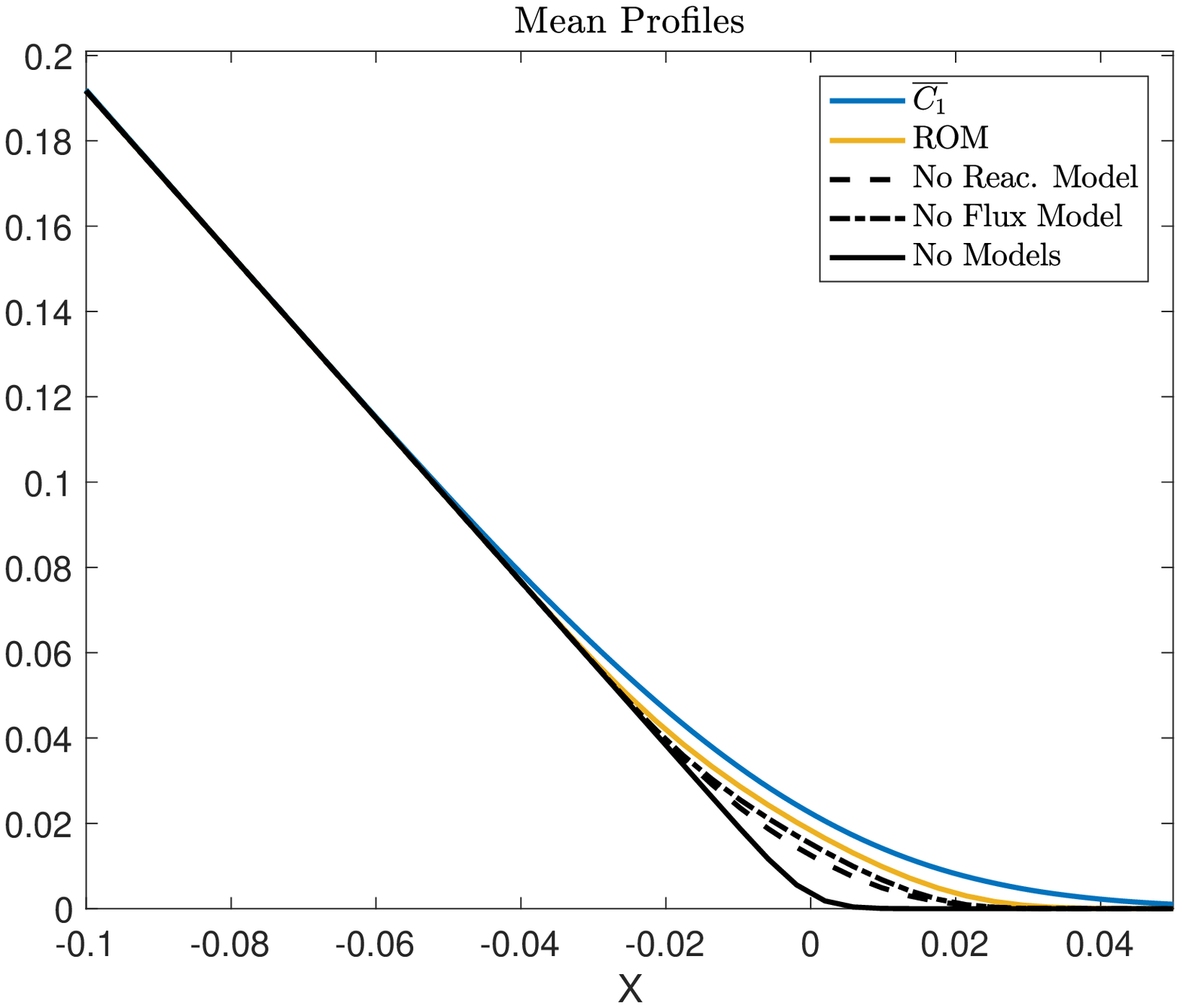}
  \caption{ }
    \label{fig:RANS_ROM}
\end{subfigure}
\caption{Results of \emph{a posteriori} analysis, with DNS quantities indicated with overbars, $X = x/L_1$. (a) A comparison of the predicted mean scalar profiles in the reaction zone for the ROM and the Gradient Diffusion models (b) A comparison of the predicted mean scalar profile for $C_1$ with different aspects of the ROM active.}
\label{fig:aposteriori}
\end{figure}

In this section, we solve Eqn. \ref{eqn:RANS_full} directly by invoking some of the closure models heretofore presented. The problem setup and parameters used in this section are identical to those for the \emph{a priori} analysis. The only exception is that for the RANS equation, instead of Dirichlet boundary conditions for the scalars, Neumann boundary conditions with slopes matching the DNS profiles are used. This is done because the DNS develops axial boundary layers near the Dirichlet conditions due to local outflow advection. For the DNS, we have ensured these artificial effects do not pollute the reaction zone results by ensuring the boundaries are far from the reaction zone. These boundary layers are absent in the RANS case as the mean velocity is zero. Appropriate matching of RANS solutions to the DNS ones should consider concentration profiles outside of these artificial boundary layers. We have done so by matching the slopes of the RANS concentration profiles to those of the DNS outside of the boundary layers, but far from the reaction zones.

In Fig. \ref{fig:RANS_full}, we compare the predicted mean profiles of $C_1$ and $C_2$ to the results derived from the DNS described in the previous section. It is obvious that when all closures are considered, the ROM outperforms the standard Gradient Diffusion model. 

To demonstrate the relative importance of each of the closure terms, Fig.\ref{fig:RANS_ROM} shows the effects of toggling each of the closure terms on the overall accuracy of scalar concentration predictions. 

\section{Discussion and Conclusions}

In this work, we have introduced a framework to extend the analysis of the dispersion of a single passive tracer to analysis of system of species undergoing binary reactions. The resulting model form derived herein introduces closures to both advective flux and reaction terms with nonlinear interactions between the mean state of all species, even though the primary set of equations being solved undergo a linearization procedure. The proposed framework also captures the non-reactive and linear reaction limits as special cases of the binary reaction problem. In contrast to \cite{battiato_2011}, the derived model leads to a positive-definite diffusion operator across all Damk\"ohler numbers, so numerical stability is maintained when the introduced ROM is used in a one-dimensional scalar equation solver.

By considering four different test cases, we showed that the standard Gradient Diffusion and Linear Reaction model do not suitably capture the behaviour of mean scalar fields across all problems. The ROM, however, can address all these different regimes consistently. Perhaps the most novel aspect of this work is the direct translation of the obtained model form between the laminar prototype problem and the test problem involving turbulent flows. Most critically, no tuning of parameters was used in this analysis, as methods like MFM permit the measurement of $D^0$ directly for turbulent flows. However, it is unclear whether omitting a constant prefactor in $\tau_{mix} = D^0/u_{rms}^2$ is appropriate for turbulent settings. In the laminar case, this definition with no need for an additional prefactor was derived analytically. In the turbulent case, we simply adopted the laminar definition without retuning its prefactor that could otherwise account for the change in flow topology.

An interesting feature of the developed ROM is that advective fluxes of any one reactant depend on mean gradients of other species involved in the reaction. This leads to an effective diffusivity matrix that shows cross-diffusion coefficients. In this sense, our results are consistent with the recent work of \cite{prend_2021} in which a two-dimensional system undergoing a slightly different reaction mechanism was considered. However, the additional advantage of the recent ROM is that it consistently models all multiphysics effects by offering closure expressions to both reaction terms as well as transport terms. Furthermore, the present ROM provides analytic model forms that capture explicitly the influence of reaction constants on each closure operator. To this end, only other inputs needed are the non-reactive eddy diffusivities and the mixing time scale to form the appropriate Damk\"ohler numbers.


It should be noted that this analysis is not identical to reacting flow formulations that consider just a mixture fraction or related quantity, as in \cite{wang_2015}, as their scalar. The mixture fraction maps both concentration fields in a binary system to a single scalar that is conserved in space. This is in contrast to the partial differential equations considered here. 


Another promising avenue of further inquiry is applying the proposed ROM to a Large-Eddy Simulation context as a subgrid-scale modes, where mean space variables are replaced with filtered variables and quantities like $u_{rms}$ are evaluated at the grid scale. Additionally, future work may be focused on capturing non-locality, as in \cite{shirian_2020}, which should account for the discrepancy of the ROM values and the DNS values. 




\section{Acknowledgements}

Support for this work was provided by the National Science Foundation Graduate Research Fellowship Program, the Stanford Graduate Fellowships in Science and Engineering, and National Science Foundation Extreme Science and Engineering Discovery Environment resources.



\bibliography{prf}

\begin{thebibliography}{21}%
\makeatletter
\providecommand \@ifxundefined [1]{%
 \@ifx{#1\undefined}
}%
\providecommand \@ifnum [1]{%
 \ifnum #1\expandafter \@firstoftwo
 \else \expandafter \@secondoftwo
 \fi
}%
\providecommand \@ifx [1]{%
 \ifx #1\expandafter \@firstoftwo
 \else \expandafter \@secondoftwo
 \fi
}%
\providecommand \natexlab [1]{#1}%
\providecommand \enquote  [1]{``#1''}%
\providecommand \bibnamefont  [1]{#1}%
\providecommand \bibfnamefont [1]{#1}%
\providecommand \citenamefont [1]{#1}%
\providecommand \href@noop [0]{\@secondoftwo}%
\providecommand \href [0]{\begingroup \@sanitize@url \@href}%
\providecommand \@href[1]{\@@startlink{#1}\@@href}%
\providecommand \@@href[1]{\endgroup#1\@@endlink}%
\providecommand \@sanitize@url [0]{\catcode `\\12\catcode `\$12\catcode
  `\&12\catcode `\#12\catcode `\^12\catcode `\_12\catcode `\%12\relax}%
\providecommand \@@startlink[1]{}%
\providecommand \@@endlink[0]{}%
\providecommand \url  [0]{\begingroup\@sanitize@url \@url }%
\providecommand \@url [1]{\endgroup\@href {#1}{\urlprefix }}%
\providecommand \urlprefix  [0]{URL }%
\providecommand \Eprint [0]{\href }%
\providecommand \doibase [0]{https://doi.org/}%
\providecommand \selectlanguage [0]{\@gobble}%
\providecommand \bibinfo  [0]{\@secondoftwo}%
\providecommand \bibfield  [0]{\@secondoftwo}%
\providecommand \translation [1]{[#1]}%
\providecommand \BibitemOpen [0]{}%
\providecommand \bibitemStop [0]{}%
\providecommand \bibitemNoStop [0]{.\EOS\space}%
\providecommand \EOS [0]{\spacefactor3000\relax}%
\providecommand \BibitemShut  [1]{\csname bibitem#1\endcsname}%
\let\auto@bib@innerbib\@empty
\bibitem [{\citenamefont {Avrin}(1997)}]{arvin_1997}%
  \BibitemOpen
  \bibfield  {author} {\bibinfo {author} {\bibfnamefont {J.~D.}\ \bibnamefont
  {Avrin}},\ }\bibfield  {title} {\bibinfo {title} {Incompressible reacting
  flows},\ }\href {http://www.jstor.org/stable/2155566} {\bibfield  {journal}
  {\bibinfo  {journal} {Transactions of the American Mathematical Society}\
  }\textbf {\bibinfo {volume} {349}},\ \bibinfo {pages} {3875} (\bibinfo {year}
  {1997})}\BibitemShut {NoStop}%
\bibitem [{\citenamefont {Peters}(2000)}]{peters_2000}%
  \BibitemOpen
  \bibfield  {author} {\bibinfo {author} {\bibfnamefont {N.}~\bibnamefont
  {Peters}},\ }\href {https://doi.org/10.1017/CBO9780511612701} {\emph
  {\bibinfo {title} {Turbulent Combustion}}},\ Cambridge Monographs on
  Mechanics\ (\bibinfo  {publisher} {Cambridge University Press},\ \bibinfo
  {year} {2000})\BibitemShut {NoStop}%
\bibitem [{\citenamefont {Seinfeld}\ and\ \citenamefont
  {Pandis}(2016)}]{seinfeld_2016}%
  \BibitemOpen
  \bibfield  {author} {\bibinfo {author} {\bibfnamefont {J.~H.}\ \bibnamefont
  {Seinfeld}}\ and\ \bibinfo {author} {\bibfnamefont {S.~N.}\ \bibnamefont
  {Pandis}},\ }\href@noop {} {\emph {\bibinfo {title} {Atmospheric chemistry
  and physics : from air pollution to climate change}}},\ \bibinfo {edition}
  {third edition.}\ ed.\ (\bibinfo  {publisher} {John Wiley and Sons},\
  \bibinfo {address} {Hoboken, New Jersey},\ \bibinfo {year}
  {2016})\BibitemShut {NoStop}%
\bibitem [{\citenamefont {Rosales}\ and\ \citenamefont
  {Nava}(2017)}]{Rosales_2017}%
  \BibitemOpen
  \bibfield  {author} {\bibinfo {author} {\bibfnamefont {M.}~\bibnamefont
  {Rosales}}\ and\ \bibinfo {author} {\bibfnamefont {J.~L.}\ \bibnamefont
  {Nava}},\ }\bibfield  {title} {\bibinfo {title} {Simulations of turbulent
  flow, mass transport, and tertiary current distribution on the cathode of a
  rotating cylinder electrode reactor in continuous operation mode during
  silver deposition},\ }\href {https://doi.org/10.1149/2.0351711jes} {\bibfield
   {journal} {\bibinfo  {journal} {Journal of The Electrochemical Society}\
  }\textbf {\bibinfo {volume} {164}},\ \bibinfo {pages} {E3345} (\bibinfo
  {year} {2017})}\BibitemShut {NoStop}%
\bibitem [{\citenamefont {Kraichnan}(1987)}]{kraichnan_1987}%
  \BibitemOpen
  \bibfield  {author} {\bibinfo {author} {\bibfnamefont {R.~H.}\ \bibnamefont
  {Kraichnan}},\ }\bibfield  {title} {\bibinfo {title} {Eddy viscosity and
  diffusivity: exact formulas and approximations},\ }\href@noop {} {\bibfield
  {journal} {\bibinfo  {journal} {Complex Systems}\ }\textbf {\bibinfo {volume}
  {1}},\ \bibinfo {pages} {805} (\bibinfo {year} {1987})}\BibitemShut {NoStop}%
\bibitem [{\citenamefont {Mani}\ and\ \citenamefont {Park}(2021)}]{mani_2021}%
  \BibitemOpen
  \bibfield  {author} {\bibinfo {author} {\bibfnamefont {A.}~\bibnamefont
  {Mani}}\ and\ \bibinfo {author} {\bibfnamefont {D.}~\bibnamefont {Park}},\
  }\bibfield  {title} {\bibinfo {title} {Macroscopic forcing method: A tool for
  turbulence modeling and analysis of closures},\ }\href
  {https://doi.org/10.1103/PhysRevFluids.6.054607} {\bibfield  {journal}
  {\bibinfo  {journal} {Phys. Rev. Fluids}\ }\textbf {\bibinfo {volume} {6}},\
  \bibinfo {pages} {054607} (\bibinfo {year} {2021})}\BibitemShut {NoStop}%
\bibitem [{\citenamefont {Taylor}(1953)}]{taylor_1953}%
  \BibitemOpen
  \bibfield  {author} {\bibinfo {author} {\bibfnamefont {G.~I.}\ \bibnamefont
  {Taylor}},\ }\bibfield  {title} {\bibinfo {title} {Dispersion of soluble
  matter in solvent flowing slowly through a tube},\ }\href@noop {} {\bibfield
  {journal} {\bibinfo  {journal} {Proceedings of the Royal Society of London.
  Series A. Mathematical and Physical Sciences}\ }\textbf {\bibinfo {volume}
  {219}},\ \bibinfo {pages} {186} (\bibinfo {year} {1953})}\BibitemShut
  {NoStop}%
\bibitem [{\citenamefont {Taylor}(1954)}]{taylor_1954}%
  \BibitemOpen
  \bibfield  {author} {\bibinfo {author} {\bibfnamefont {G.~I.}\ \bibnamefont
  {Taylor}},\ }\bibfield  {title} {\bibinfo {title} {The dispersion of matter
  in turbulent flow through a pipe},\ }\href
  {https://doi.org/10.1098/rspa.1954.0130} {\bibfield  {journal} {\bibinfo
  {journal} {Proceedings of the Royal Society of London. Series A. Mathematical
  and Physical Sciences}\ }\textbf {\bibinfo {volume} {223}},\ \bibinfo {pages}
  {446} (\bibinfo {year} {1954})}\BibitemShut {NoStop}%
\bibitem [{\citenamefont {Aris}(1956)}]{aris_1956}%
  \BibitemOpen
  \bibfield  {author} {\bibinfo {author} {\bibfnamefont {R.}~\bibnamefont
  {Aris}},\ }\bibfield  {title} {\bibinfo {title} {On the dispersion of a
  solute in a fluid flowing through a tube},\ }\href@noop {} {\bibfield
  {journal} {\bibinfo  {journal} {Proceedings of the Royal Society of London.
  Series A. Mathematical and Physical Sciences}\ }\textbf {\bibinfo {volume}
  {235}},\ \bibinfo {pages} {67 } (\bibinfo {year} {1956})}\BibitemShut
  {NoStop}%
\bibitem [{\citenamefont {Lighthill}(1966)}]{lighthill_1966}%
  \BibitemOpen
  \bibfield  {author} {\bibinfo {author} {\bibfnamefont {M.~J.}\ \bibnamefont
  {Lighthill}},\ }\bibfield  {title} {\bibinfo {title} {{Initial Development of
  Diffusion in Poiseuille Flow}},\ }\href
  {https://doi.org/10.1093/imamat/2.1.97} {\bibfield  {journal} {\bibinfo
  {journal} {IMA Journal of Applied Mathematics}\ }\textbf {\bibinfo {volume}
  {2}},\ \bibinfo {pages} {97} (\bibinfo {year} {1966})}\BibitemShut {NoStop}%
\bibitem [{\citenamefont {Shapiro}\ and\ \citenamefont
  {Brenner}(1986)}]{shapiro_1986}%
  \BibitemOpen
  \bibfield  {author} {\bibinfo {author} {\bibfnamefont {M.}~\bibnamefont
  {Shapiro}}\ and\ \bibinfo {author} {\bibfnamefont {H.}~\bibnamefont
  {Brenner}},\ }\bibfield  {title} {\bibinfo {title} {Taylor dispersion of
  chemically reactive species: Irreversible first-order reactions in bulk and
  on boundaries},\ }\href
  {https://doi.org/https://doi.org/10.1016/0009-2509(86)85228-9} {\bibfield
  {journal} {\bibinfo  {journal} {Chemical Engineering Science}\ }\textbf
  {\bibinfo {volume} {41}},\ \bibinfo {pages} {1417} (\bibinfo {year}
  {1986})}\BibitemShut {NoStop}%
\bibitem [{\citenamefont {Corrsin}(1961)}]{corrsin_1961}%
  \BibitemOpen
  \bibfield  {author} {\bibinfo {author} {\bibfnamefont {S.}~\bibnamefont
  {Corrsin}},\ }\bibfield  {title} {\bibinfo {title} {The reactant
  concentration spectrum in turbulent mixing with a first-order reaction},\
  }\href {https://doi.org/10.1017/S0022112061000615} {\bibfield  {journal}
  {\bibinfo  {journal} {Journal of Fluid Mechanics}\ }\textbf {\bibinfo
  {volume} {11}},\ \bibinfo {pages} {407–416} (\bibinfo {year}
  {1961})}\BibitemShut {NoStop}%
\bibitem [{\citenamefont {Battiato}\ and\ \citenamefont
  {Tartakovsky}(2011)}]{battiato_2011}%
  \BibitemOpen
  \bibfield  {author} {\bibinfo {author} {\bibfnamefont {I.}~\bibnamefont
  {Battiato}}\ and\ \bibinfo {author} {\bibfnamefont {D.}~\bibnamefont
  {Tartakovsky}},\ }\bibfield  {title} {\bibinfo {title} {Applicability regimes
  for macroscopic models of reactive transport in porous media},\ }\href
  {https://doi.org/https://doi.org/10.1016/j.jconhyd.2010.05.005} {\bibfield
  {journal} {\bibinfo  {journal} {Journal of Contaminant Hydrology}\ }\textbf
  {\bibinfo {volume} {120-121}},\ \bibinfo {pages} {18} (\bibinfo {year}
  {2011})}\BibitemShut {NoStop}%
\bibitem [{\citenamefont {Elperin}\ \emph {et~al.}(2014)\citenamefont
  {Elperin}, \citenamefont {Kleeorin}, \citenamefont {Liberman},\ and\
  \citenamefont {Rogachevskii}}]{elperin_2014}%
  \BibitemOpen
  \bibfield  {author} {\bibinfo {author} {\bibfnamefont {T.}~\bibnamefont
  {Elperin}}, \bibinfo {author} {\bibfnamefont {N.}~\bibnamefont {Kleeorin}},
  \bibinfo {author} {\bibfnamefont {M.}~\bibnamefont {Liberman}},\ and\
  \bibinfo {author} {\bibfnamefont {I.}~\bibnamefont {Rogachevskii}},\
  }\bibfield  {title} {\bibinfo {title} {Turbulent diffusion of chemically
  reacting gaseous admixtures},\ }\href
  {https://doi.org/10.1103/PhysRevE.90.053001} {\bibfield  {journal} {\bibinfo
  {journal} {Phys. Rev. E}\ }\textbf {\bibinfo {volume} {90}},\ \bibinfo
  {pages} {053001} (\bibinfo {year} {2014})}\BibitemShut {NoStop}%
\bibitem [{\citenamefont {Elperin}\ \emph {et~al.}(2017)\citenamefont
  {Elperin}, \citenamefont {Kleeorin}, \citenamefont {Liberman}, \citenamefont
  {Lipatnikov}, \citenamefont {Rogachevskii},\ and\ \citenamefont
  {Yu}}]{elperin_2017}%
  \BibitemOpen
  \bibfield  {author} {\bibinfo {author} {\bibfnamefont {T.}~\bibnamefont
  {Elperin}}, \bibinfo {author} {\bibfnamefont {N.}~\bibnamefont {Kleeorin}},
  \bibinfo {author} {\bibfnamefont {M.}~\bibnamefont {Liberman}}, \bibinfo
  {author} {\bibfnamefont {A.~N.}\ \bibnamefont {Lipatnikov}}, \bibinfo
  {author} {\bibfnamefont {I.}~\bibnamefont {Rogachevskii}},\ and\ \bibinfo
  {author} {\bibfnamefont {R.}~\bibnamefont {Yu}},\ }\bibfield  {title}
  {\bibinfo {title} {Turbulent diffusion of chemically reacting flows: Theory
  and numerical simulations},\ }\href
  {https://doi.org/10.1103/PhysRevE.96.053111} {\bibfield  {journal} {\bibinfo
  {journal} {Phys. Rev. E}\ }\textbf {\bibinfo {volume} {96}},\ \bibinfo
  {pages} {053111} (\bibinfo {year} {2017})}\BibitemShut {NoStop}%
\bibitem [{\citenamefont {Watanabe}\ \emph {et~al.}(2014)\citenamefont
  {Watanabe}, \citenamefont {Sakai}, \citenamefont {Nagata},\ and\
  \citenamefont {Terashima}}]{watanabe_2014}%
  \BibitemOpen
  \bibfield  {author} {\bibinfo {author} {\bibfnamefont {T.}~\bibnamefont
  {Watanabe}}, \bibinfo {author} {\bibfnamefont {Y.}~\bibnamefont {Sakai}},
  \bibinfo {author} {\bibfnamefont {K.}~\bibnamefont {Nagata}},\ and\ \bibinfo
  {author} {\bibfnamefont {O.}~\bibnamefont {Terashima}},\ }\bibfield  {title}
  {\bibinfo {title} {Turbulent schmidt number and eddy diffusivity change with
  a chemical reaction},\ }\href {https://doi.org/10.1017/jfm.2014.387}
  {\bibfield  {journal} {\bibinfo  {journal} {Journal of Fluid Mechanics}\
  }\textbf {\bibinfo {volume} {754}},\ \bibinfo {pages} {98–121} (\bibinfo
  {year} {2014})}\BibitemShut {NoStop}%
\bibitem [{\citenamefont {Shirian}\ and\ \citenamefont
  {Mani}(2019)}]{shirian_2020}%
  \BibitemOpen
  \bibfield  {author} {\bibinfo {author} {\bibfnamefont {Y.}~\bibnamefont
  {Shirian}}\ and\ \bibinfo {author} {\bibfnamefont {A.}~\bibnamefont {Mani}},\
  }\href@noop {} {\bibinfo {title} {Eddy diffusivity in homogeneous isotropic
  turbulence}} (\bibinfo {year} {2019}),\ \Eprint
  {https://arxiv.org/abs/1905.08379} {arXiv:1905.08379 [physics.flu-dyn]}
  \BibitemShut {NoStop}%
\bibitem [{\citenamefont {Pouransari}\ \emph {et~al.}(2016)\citenamefont
  {Pouransari}, \citenamefont {Mortazavi},\ and\ \citenamefont
  {Mani}}]{pouransari_2016}%
  \BibitemOpen
  \bibfield  {author} {\bibinfo {author} {\bibfnamefont {H.}~\bibnamefont
  {Pouransari}}, \bibinfo {author} {\bibfnamefont {M.}~\bibnamefont
  {Mortazavi}},\ and\ \bibinfo {author} {\bibfnamefont {A.}~\bibnamefont
  {Mani}},\ }\href {hadip@bitbucket.org/hadip/soleilmpi} {\bibinfo {title}
  {Parallel variable-density particle-laden turbulence simulation}} (\bibinfo
  {year} {2016}),\ \Eprint {https://arxiv.org/abs/1601.05448}
  {arXiv:1601.05448} \BibitemShut {NoStop}%
\bibitem [{\citenamefont {Rosales}\ and\ \citenamefont
  {Meneveau}(2005)}]{rosales_2005}%
  \BibitemOpen
  \bibfield  {author} {\bibinfo {author} {\bibfnamefont {C.}~\bibnamefont
  {Rosales}}\ and\ \bibinfo {author} {\bibfnamefont {C.}~\bibnamefont
  {Meneveau}},\ }\bibfield  {title} {\bibinfo {title} {Linear forcing in
  numerical simulations of isotropic turbulence: Physical space implementations
  and convergence properties},\ }\href {https://doi.org/10.1063/1.2047568}
  {\bibfield  {journal} {\bibinfo  {journal} {Physics of Fluids}\ }\textbf
  {\bibinfo {volume} {17}},\ \bibinfo {pages} {095106} (\bibinfo {year}
  {2005})}\BibitemShut {NoStop}%
\bibitem [{\citenamefont {Prend}\ \emph {et~al.}(2021)\citenamefont {Prend},
  \citenamefont {Flierl}, \citenamefont {Smith},\ and\ \citenamefont
  {Kaminski}}]{prend_2021}%
  \BibitemOpen
  \bibfield  {author} {\bibinfo {author} {\bibfnamefont {C.~J.}\ \bibnamefont
  {Prend}}, \bibinfo {author} {\bibfnamefont {G.~R.}\ \bibnamefont {Flierl}},
  \bibinfo {author} {\bibfnamefont {K.~M.}\ \bibnamefont {Smith}},\ and\
  \bibinfo {author} {\bibfnamefont {A.~K.}\ \bibnamefont {Kaminski}},\
  }\bibfield  {title} {\bibinfo {title} {Parameterizing eddy transport of
  biogeochemical tracers},\ }\href
  {https://doi.org/https://doi.org/10.1029/2021GL094405} {\bibfield  {journal}
  {\bibinfo  {journal} {Geophysical Research Letters}\ }\textbf {\bibinfo
  {volume} {48}},\ \bibinfo {pages} {e2021GL094405} (\bibinfo {year}
  {2021})}\BibitemShut {NoStop}%
\bibitem [{\citenamefont {Wang}\ \emph {et~al.}(2015)\citenamefont {Wang},
  \citenamefont {Sun}, \citenamefont {Yang},\ and\ \citenamefont
  {Qin}}]{wang_2015}%
  \BibitemOpen
  \bibfield  {author} {\bibinfo {author} {\bibfnamefont {H.}~\bibnamefont
  {Wang}}, \bibinfo {author} {\bibfnamefont {M.}~\bibnamefont {Sun}}, \bibinfo
  {author} {\bibfnamefont {Y.}~\bibnamefont {Yang}},\ and\ \bibinfo {author}
  {\bibfnamefont {N.}~\bibnamefont {Qin}},\ }\bibfield  {title} {\bibinfo
  {title} {A passive scalar-based method for numerical combustion},\ }\href
  {https://doi.org/https://doi.org/10.1016/j.ijhydene.2015.06.148} {\bibfield
  {journal} {\bibinfo  {journal} {International Journal of Hydrogen Energy}\
  }\textbf {\bibinfo {volume} {40}},\ \bibinfo {pages} {10658} (\bibinfo {year}
  {2015})}\BibitemShut {NoStop}%
\end{thebibliography}%

\end{document}